\documentclass{article}
\usepackage{graphicx}
\usepackage{amsfonts}
\usepackage[dvipsnames]{xcolor}
\usepackage{amsmath}
\usepackage{verbatim}
\usepackage[numbers]{natbib}
\usepackage{accents}
\usepackage{adjustbox}
\usepackage{subcaption}
\usepackage{float}
\usepackage{bbm}
\usepackage{soul}

\title{The asymptotic distribution of the likelihood ratio test statistic in two-peak discovery experiments}
\author{Clara Bertinelli Salucci, Hedvig Borgen Reiersrud, A. L. Read,\\ Anders Kvellestad, Riccardo De Bin}
\date{December 2025}

\definecolor{darkgreen}{rgb}{0.0, 0.5, 0.25}

\newcommand{\EMIN}{E_{\mbox{\tiny min}}}
\newcommand{\EMAX}{E_{\mbox{\tiny max}}}

\DeclareMathOperator{\lrs}{\lambda_{\mbox{\scriptsize LR}}}

\begin{document}

\maketitle

\section*{Abstract}

Likelihood ratio tests are widely used in high-energy physics, where the test statistic is usually assumed to follow a chi-squared distribution with a number of degrees of freedom specified by Wilks' theorem. This assumption breaks down when parameters such as signal or coupling strengths are restricted to be non-negative and their values under the null hypothesis lie on the boundary of the parameter space. Based on a recent clarification concerning the correct asymptotic distribution of the likelihood ratio test statistic for cases where two of the parameters are on the boundary, we revisit the the question of significance estimation for two-peak signal-plus-background counting experiments. In the high-energy physics literature, such experiments are commonly analyzed using Wilks' chi-squared distribution or the one-parameter Chernoff limit. We demonstrate that these approaches can lead to strongly miscalibrated significances, and that the test statistic distribution is instead well described by a chi-squared mixture with weights determined by the Fisher information matrix. Our results highlight the need for boundary-aware asymptotics in the analysis of two-peak counting experiments.

\section{Introduction}

In likelihood-based statistical inference, particularly when conducting hypothesis testing via likelihood ratio tests, it is almost universally assumed that the distribution of the test statistic under the null hypothesis can be asymptotically approximated by a chi-squared distribution with degrees of freedom equal to the number of parameters of interest under consideration, as by Wilks' theorem \cite{wilks}. It is frequently overlooked that this theorem relies upon a set of regularity conditions -- including identifiability, parameter-independent support, interiority of the true parameter, smoothness of the log-likelihood, satisfaction of Bartlett's identities -- which are not so rarely violated. Recently, renewed attention to this oversight has been drawn in the works by Brazzale and Mameli \cite{brazzale} and Algeri et al. \cite{algeri}. Violations of the interiority condition occur whenever the null hypothesis fixes a parameter at the boundary of its domain, so that the maximum likelihood estimator is restricted to one side of the boundary. As a consequence, the usual symmetry underlying classical asymptotics is broken, so that estimators and likelihood ratio statistics exhibit nonstandard limiting behaviour. However, the availability of a closed-form expression for the asymptotic distribution even in such cases is particularly valuable in practice, as simulation-based or data-driven calibration can be computationally prohibitive \cite{glen_cowan} or even infeasible.

The first rigorous analysis of this problem was provided by Chernoff \cite{chernoff}, who showed that testing a single parameter of interest on the boundary leads to a limiting distribution that is a 50:50 mixture of a point mass at zero and a $\chi_1^2$, often written as $\tfrac12 \chi_0^2 + \tfrac12 \chi_1^2$. When extending to more than one constrained parameter, it is important to distinguish between parameters of interest and nuisance parameters, the latter being typically handled through likelihood profiling. In the case of two parameters of interest on the boundary, the null distribution of the likelihood ratio statistic is a mixture of $\chi_0^2$, $\chi_1^2$, and $\chi_2^2$, with weights determined by the Fisher information matrix at the boundary \cite{SL, us}. When the two parameters on the boundary are a parameter of interest and a nuisance parameter, if they are positively correlated\footnote{Throughout, when referring to parameters as being correlated, we mean that their estimators are correlated, as quantified by the covariance (or Fisher information) matrix in a frequentist framework.}, we still have the same mixture of $\chi_0^2$, $\chi_1^2$, and $\chi_2^2$ but with weights that differ in form and are governed by a different functional of the Fisher information \cite{KS}. In the case of negative correlation, the null distribution is no longer exactly a chi-squared mixture, but it can be closely approximated by a modified version of the mixture valid under positive correlation \cite{us}.

In the physics literature, however, it is common to rely on Wilks' chi-squared distribution, at most correcting for a single parameter on the boundary. Examples include line searches in $\gamma$-ray astronomy, which often report local significances using Wilks' or Chernoff's distribution, also while scanning models that actually involve two constrained amplitudes (double-line searches) \cite{FermiLAT, anderson_2016, rev_2}, or perform per-line tests while treating the second line's amplitude as a nuisance parameter \cite{fermilat1}, overlooking that, even as a nuisance, the parameter still lies on the boundary and must be accounted for accordingly. Collider spectroscopy analyses, too, typically test one peak at a time or compare one-peak vs two-peak signal-plus-background models within extended-likelihood fits to invariant-mass spectra \cite{cms, lhcb}.  This practice is adequate for establishing the presence of individual structures, but when a well-motivated theory links two nearby peaks, e.g.\ by fixing their relative position or strength, joint testing can increase power, sharpen intervals, and avoid double-counting look-elsewhere effects that inevitably arise from separate per-peak scans. 

One may wonder whether adopting a different asymptotic distribution is really necessary, since the null hypothesis in both single-peak and double-peak searches is usually still ``no peaks at all''. The key point is that the structure of the alternative hypothesis matters. Allowing two possible non-zero amplitudes enlarges the parameter space, and the likelihood ratio test must accommodate the possibility that the data may favour either or both peaks. Even under background-only data, the maximum likelihood estimator may therefore lie on one or both boundaries, which alters the geometry of the problem and changes the limiting distribution of the test statistic, even though the null hypothesis itself is unchanged. A related scenario arises when treating one of the two amplitudes as the 
parameter of interest and the second as a nuisance, as often done in the 
literature. However, under background-only data, the nuisance 
amplitude remains on the boundary, and its presence modifies the asymptotic law of the likelihood ratio statistic, though in a different way from the case in which both amplitudes are treated as parameters of interest. As we show here, using the incorrect asymptotic reference leads to a systematic miscalibration of $p$-values and local significances, with direct consequences for the reliability of discovery claims.

In this article, we adopt a paradigmatic toy example to demonstrate through physics-motivated simulations the correct asymptotic distribution for the likelihood ratio test statistic when two parameters are on the boundary. We consider a signal-plus-background experiment akin to the one described in \cite{algeri}, but with a double peak signal, which provides a good abstraction of the scenarios involved in the aforementioned articles. We examine the consequences of relying on incorrect reference distributions when the two parameters on the boundary are both of interest, and when one of them is a nuisance. Particular attention is given to the choice of parametrization, which proves critical for maintaining the validity of large-sample theory. We also derive and present the analytic expressions for the Fisher information matrix that determines the geometry of the problem and the weights of the resulting chi-squared mixture distribution.

The remainder of this article is organized as follows: In Section \ref{sec:case_study} we describe the two-peak signal-plus-background case. In Section \ref{sec:lrt} we introduce the likelihood ratio test and asymptotic results for one and two parameters on the boundary, discuss the appropriate parametrization for the case under consideration and present the simulation results. The implications of these findings are analyzed in Section \ref{sec:discussion}, where we also comment on the applicability of the framework across different experimental scenarios. In Section \ref{sec:conclusion} we present our conclusions.

\section{Two-peak signal-plus-background experiment} \label{sec:case_study}

\begin{figure}[H]
  \centering
       \includegraphics[width=.6\linewidth]{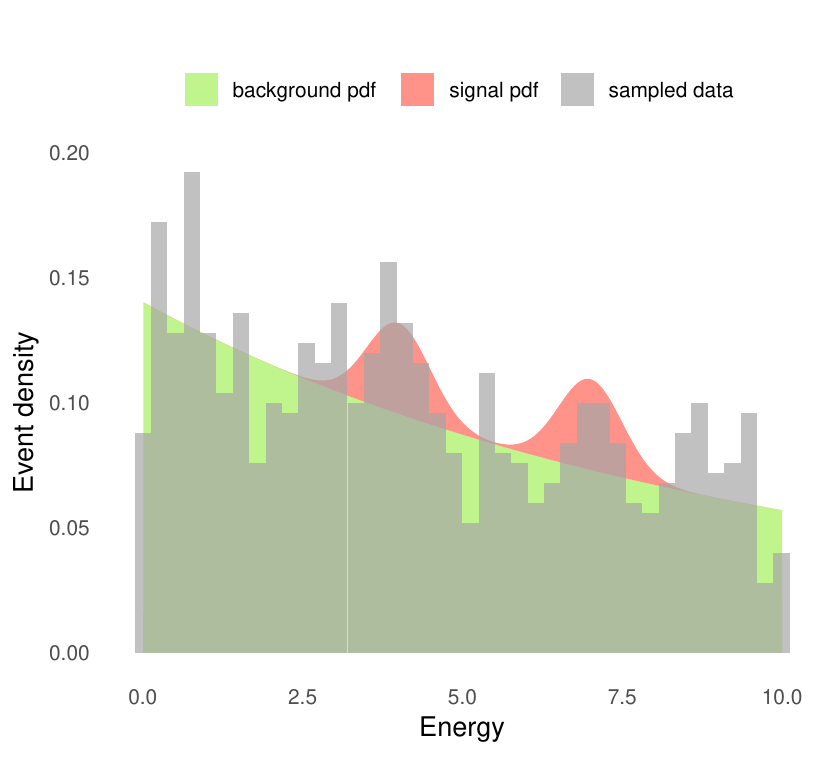}
  \caption{\small Illustration of the two-peak toy model used throughout the analysis. The green area corresponds to the background component, modelled as a normalized exponential density; the red area represents the sum of two truncated Gaussian signal components (double-peak signal); and the grey histogram shows a realization of data simulated under the null hypothesis.}
  \label{fig:stacked_true_densities}
\end{figure}

Consider an experiment in which $N$ events are observed within an energy range $[E_{\mbox{\tiny min}},E_{\mbox{\tiny max}}]$, each event $i$ ($i = 1,2,\ldots,N$) being characterized by its measured energy $E_i$. The analysis aims to investigate the possible presence of a new phenomenon producing two distinct structures in the spectrum, quantified by the mean numbers $\mu_1$ and $\mu_2$ of signal events, occurring on top of a background with expected yield $\beta$. The signals are modelled as Gaussian densities truncated to the energy interval, with known means $\gamma_1$ and $\gamma_2$ and unit variance,
\[
g_j(E) \;=\; \frac{\exp\!\left(-\tfrac{1}{2}(E-\gamma_j)^2\right)}{\int_{\EMIN}^{\EMAX} \exp\!\left(-\tfrac{1}{2}(u-\gamma_j)^2\right)\,du}, 
\qquad j=1,2.
\]
The background is described by an exponential density,  
\[
b(E; \phi_1, \phi_2) = \frac{\exp(\phi_1 E + \phi_2 E^2)}{\int_{\EMIN}^{\EMAX} \exp(\phi_1 u + \phi_2 u^2)\,du},
\]  
and both signal and background densities are normalized to integrate to unity over the considered interval. Figure~\ref{fig:stacked_true_densities} displays an example of this configuration, where the two signal peaks are superimposed on the exponential background, together with observations sampled under the null hypothesis (background only). This toy example is essentially a two-peak extension of the example considered in \cite{algeri}. It is meant as a paradigmatic abstraction of many problems of this kind, and, although it directly extends the one-peak setting, the presence of two parameters on the boundary rather than one will require a substantially different analysis. 

The expected signal counts $\mu_1$ and $\mu_2$ are the parameters of interest, and the expected background counts $\beta$, along with the shape parameters $\phi_1$ and $\phi_2$ of the background model, are treated as nuisance parameters, i.e.\ parameters not of direct interest but whose estimation is required to perform inference. The null hypothesis corresponds to the absence of both signals, $\mu_1 = \mu_2 = 0$, while the alternative hypothesis entails nonzero $\mu_1$ and/or $\mu_2$. Since the signal counts cannot be negative, setting \(\mu_1 = \mu_2 = 0\) under the null places both parameters of interest on the boundary of the parameter space. 

Let \(\lambda = \mu_1 + \mu_2 + \beta\) denote the total expected number of events. Given a sample of $N$ observed energies $E_1, \dots, E_N$, the extended likelihood function for $\theta = (\mu_1, \mu_2, \beta, \phi_1, \phi_2)$ is usually expressed as \cite{algeri}

\begin{equation}
L(\theta) = \text{Pois}(N \, \vert \, \lambda) \cdot \prod_{i=1}^{N}  \frac{1}{\lambda} \, \left[ \,
\mu_1 \, g_1(E_i) +
\mu_2 \, g_2(E_i) +
\beta \, b(E_i; \phi_1, \phi_2)
\, \right], 
\label{likelihood}
\end{equation}

\noindent with log-likelihood

\begin{equation}
\ell(\theta) = \sum_{i=1}^{N} \log \left[ \frac{1}{\lambda} \left( 
\mu_1 \, g_1(E_i) +
\mu_2 \, g_2(E_i) +
\beta \, b(E_i; \phi_1, \phi_2) \right)
\right] 
- \lambda + N \log\lambda - \log(N!),
\label{logl}
\end{equation}

\noindent where the Poisson probability mass function $\text{Pois}(N \, \vert \, \lambda) = \frac{\lambda^N e^{-\lambda}}{N!}$ accounts for the fact that the total number of observed events \( N \) is itself a stochastic quantity. As shown in the next section, the extended likelihood function under this parametrization fails to satisfy the conditions of classical asymptotic theory, hence requiring an alternative parametrization.

\section{Asymptotic results for the likelihood ratio test statistic} \label{sec:lrt}

The likelihood ratio test is one of the most widely used tools for hypothesis testing in experimental physics, owing to its generality and its asymptotic properties \cite{casella2002, cowan1998}. Its fundamental role is grounded in the Neyman–Pearson lemma, which establishes the likelihood ratio test as the most powerful test for discriminating between two simple hypotheses \cite{neyman-pearson}. Under regular conditions, Wilks' theorem guarantees that the distribution of the likelihood ratio test statistic, $\lrs$,  converges under the null hypothesis to a chi-squared distribution with degrees of freedom equal to the number of parameters being tested. This provides the theoretical basis for the standard practice of translating the test statistic into $p$-values or local significances. Formally, given the likelihood $L(\theta)$ based on a sample of size $N$, $\lrs$ is defined as

\[
  \lrs \;=\;-2\log \left( \frac{\sup_{\theta\in \Theta_0}L(\theta)}{\sup_{\theta\in \Theta_1}L(\theta)} \right).
\]

\noindent where $\theta$ is a vector of parameters, and $\Theta_0$ and $\Theta_1$ denote the parameter spaces under the null and alternative hypotheses, respectively. By construction, large values of $\lrs$ provide evidence against the null hypothesis in favour of the alternative. 

In this setting, we cannot rely directly on Wilks' theorem, since the presence of parameters on the boundary violates its regularity conditions \cite{brazzale, algeri}. However, we can consider alternative asymptotic chi-squared distributions that account for boundary parameters \citep[][]{us}, provided that the remaining regularity conditions are satisfied. For example, in this analysis the means of the signal Gaussian peaks are treated as known, because they would otherwise become unidentifiable under the null hypothesis, thereby violating regularity. Another key requirement is the correct specification of the asymptotic framework, ensuring that $N \rightarrow \infty$ genuinely corresponds to the addition of statistical units, ensuring a proper specification of th asymptotic framework.

\subsection{Parametrization and asymptotic regime} \label{sec:parametrization}

When the extended-likelihood formalism is written as in Eq.~(\ref{likelihood}) -- 
parameterized in terms of the expected \emph{counts} $\mu$'s and $\beta$ and
expressed as a product over the $N$ detected events -- one might be tempted to regard each
observed event as the fundamental observational unit, and to imagine the asymptotic regime to be reached as $N \to \infty$.  This interpretation mirrors familiar likelihood constructions for Gaussian, exponential, or binomial samples, where
each measurement indeed constitutes an independent observational unit.  

However, this analogy is misleading for Poisson-distributed data: in a counting
experiment, the Poisson count does not arise from an underlying sequence of independent observational units analogous to individual measurements. Rather, the observational unit is the number of observed events $N$ for specific $\mu$'s and $\beta$ and for a well-specified exposure $T$, i.e., the natural observational unit is the \emph{exposure interval} over which the Poisson process is observed.  As shown in Appendix A, indeed, if one treats the single event as the observational unit, Bartlett's identities fail, indicating that the parameterization in Eq.~(\ref{likelihood}) does not correspond to a proper likelihood for any well-defined sampling unit. 

To make this explicit, consider the Fisher information matrix,
\[
\mathcal{I}(\theta) = -\,\mathbb{E}_\theta\!\left[ \nabla^2 \ell(\theta) \right],
\]
which, under regularity conditions, characterizes the asymptotic covariance of
the maximum-likelihood estimator (MLE). Here, for simplicity, we restrict attention to a single signal peak with strength $\mu$, normalized density $g(E)$, and null hypothesis $H_0 : \mu = 0$; the extension to a two-peak model follows directly. Under the parametrization of
Eq.~(\ref{likelihood}), increasing the sample size under the null hypothesis
necessarily corresponds to increasing the expected background $\beta$.  As shown
in the appendix, the Fisher information for $\mu$ takes the form
\[
\mathcal{I}_{\mu}(\theta_0)
   = \frac{1}{\beta_0}
     \int_{\EMIN}^{\EMAX}
     \frac{g(E)^2}{b(E)} \, dE,
\]
which vanishes as $\beta_0 \to \infty$, where $\beta_0$ denotes the true value of the expected background count $\beta$.  Thus, in this parametrization, the
large-sample limit paradoxically removes information: the variance of any
unbiased estimator of $\mu$ diverges, and the likelihood surface becomes
asymptotically flat in the $\mu$-direction.  Classical asymptotic theory breaks
down, as the estimator of $\mu$ becomes increasingly unreliable for increasing $\beta_0$.

A statistically consistent formulation is obtained by treating the exposure interval as the observational unit and modelling the experiment as a Poisson point process (PPP) observed over intervals of length $T$.  In this formulation,
$\mu$ and $\beta$ represent \emph{rates} (events per unit exposure), and the
expected numbers of signal and background events in an interval are $\mu T$ and
$\beta T$.  Letting $h(E_i) = \mu g(E_i) + \beta b(E_i)$ for $i = 1,...,N$, the likelihood for a single exposure interval is
\begin{equation}
L_T(\theta)
   = \frac{e^{-(\mu + \beta)T}\,[(\mu+\beta)T]^N}{N!}
     \prod_{i=1}^N \frac{h(E_i)}{\mu+\beta},
\label{exp-likelihood}
\end{equation}
with log-likelihood
\[
\ell_T(\theta)
   = -(\mu + \beta)T + N \log T
     + \sum_{i=1}^N \log h(E_i) - \log N!.
\]

\noindent In this parametrization the asymptotic regime corresponds to increasing the exposure, not the rates, and the Fisher information grows linearly with total
exposure $T_{\mathrm{tot}} = nT$. In fact, letting $\mathcal{I}_\mu(\theta_0; T)$ denote the Fisher information associated with a single exposure interval of duration $T$, we obtain
\[
n\,\mathcal{I}_\mu(\theta_0; T)
   = T_{\mathrm{tot}} \cdot I_\mu^{(1)}(\theta_0),
   \qquad
   I_\mu^{(1)}(\theta_0)
   = \frac{1}{\beta_0}
     \int_{\EMIN}^{\EMAX}
     \frac{g(E)^2}{b(E)}\,dE,
\]
hence
\[
\mathcal{I}_\mu(\theta_0; T_{\mathrm{tot}})
   = \frac{T_{\mathrm{tot}}}{\beta_0}
     \int_{\EMIN}^{\EMAX}
     \frac{g(E)^2}{b(E)}\,dE.
\]
Therefore, the Fisher information never vanishes in the large-exposure limit and
Bartlett's identities hold for any number of exposure intervals.  

\smallskip

Note that this parametrization issue is not specific to the extended likelihood, but arises in any Poisson likelihood used for counting experiments whenever the observational unit is implicitly taken to be the detected event rather than the exposure. While in many analyses a parametrization that implicitly treats the exposure as ``built in'' to the absolute counts works well in practice, yielding stable point estimates and reasonable variance estimates, the asymptotic interpretation of the likelihood is nonetheless incorrect. This becomes important in situations where the Fisher information plays a direct role in the limiting distribution of test statistic, such as the two-peak discovery test statistic considered in this article; however, the explicit PPP parametrization is also in general required to obtain a well-defined asymptotic regime. Otherwise, even though practical difficulties may not arise, especially when focusing on point estimation \citep{lyons}, the theoretical properties of the likelihood-ratio statistic -- such as efficiency, optimality, and the validity of its limiting distribution -- are no longer guaranteed.

\subsection{Boundaries and asymptotic distribution of the likelihood ratio test statistic} 

While the correct asymptotic regime can be recovered by an appropriate reparameterization of the problem, there is no chance to correctly invoke Wilks' theorem when the assumption that the true parameter lies in the interior of the parameter space is violated. This issue is known in the physics literature, and the correct asymptotic distribution for a single parameter of interest on the boundary, i.e.\ the 50:50 mixture $\tfrac12 \chi_0^2 + \tfrac12 \chi_1^2$ \cite{chernoff, glen_cowan}, is correctly used (e.g.\ \cite{cms_chernoff, bringmann_hatlen}). It is worth stressing, however, that this result is only valid for the simple case of one parameter on the boundary, and does not generalize to the two-parameter case, even if one of the two parameters on the boundary is a nuisance parameter.

\subsubsection{Two parameters of interest on the boundary}
For two parameters of interest on the boundary, the distribution is a mixture of $\chi_0^2$, $\chi_1^2$ and $\chi_2^2$ with weights 

\[
\{w_0, w_1, w_2\} \;\;=\;\; \left\{ \dfrac12 - \frac{\arccos\left(\frac{I_{12}}{\sqrt{I_{11}I_{22}}}\right)}{2\pi}, \; \; \dfrac12, \; \; \frac{\arccos\left(\frac{I_{12}}{\sqrt{I_{11}I_{22}}}\right)}{2\pi}  \right\},
\]

\noindent where $I = I(\theta_0)$ denotes the expected Fisher information matrix evaluated at the true parameter values $\theta_0$ \cite{SL, us}. 
The entries $I_{11}$ and $I_{22}$ correspond to the two parameters of interest, while $I_{12}$ is their off-diagonal element, which captures the correlation structure. Since the true parameter value is unknown, in practice it is often replaced by the maximum likelihood estimate $\hat{\theta}$. In connection with the discussion in Section~\ref{sec:parametrization}, we note that both Chernoff's result for one parameter on the boundary and the extension by Self and Liang for two parameters on the boundary explicitly require the Fisher information matrix to be positive definite at the true parameter value: in other words, for these results to hold, the Fisher information cannot be vanishing. 

\subsubsection{A nuisance parameter and a parameter of interest on the boundary}
When both a nuisance parameter and a parameter of interest lie on the boundary, the asymptotic distribution of $\lrs$ remains a mixture of $\chi^2_0$, $\chi^2_1$, and $\chi^2_2$ in the case of positive correlation, but the mixing weights differ \cite{KS, us}: 

\begin{equation}
\{\tilde{w}_0, \tilde{w}_1, \tilde{w}_2\} \;\;=\;\; \left\{ \dfrac12 - \frac{\arcsin\left(-\frac{I_{12}}{\sqrt{I_{11}I_{22}}}\right)}{2\pi}, \; \; \dfrac12, \; \; \frac{\arcsin\left(-\frac{I_{12}}{\sqrt{I_{11}I_{22}}}\right)}{2\pi}  \right\}.
\label{weights_1001}
\end{equation}

\noindent When the correlation between the boundary parameters is negative, i.e.\ $\rho = -\tfrac{I_{12}}{\sqrt{I_{11}I_{22}}} < 0$, the asymptotic distribution of $\lrs$ can no longer be expressed as a chi-squared mixture. Nevertheless, \cite{us} shows that the mixture valid for $\rho > 0$ can be suitably modified to provide an accurate approximation also in the negatively correlated case,

\begin{equation*}
    f_{\mbox{\tiny mod}}(x) \;=\; 
    \tfrac{1}{2} f_{\chi_0^2}(x) \, \\ \;+\; \tfrac{1}{2} \, f_{\chi^2_1}(x) 
    \;+\; \tilde{w}_2 \, f_{\chi^2_2}(x) \;-\; \tilde{w}_2 \, g_\varepsilon(x), 
    \qquad \tilde{w}_2 = \frac{\arcsin(\rho)}{2\pi} < 0 ,
\end{equation*}

\noindent where $g_\varepsilon(x)$ denotes the pdf of any non-negative distribution supported on $(0,\varepsilon)$ and normalized to integrate to one, for example the uniform density on $(0,\varepsilon)$; in practice, small values of $\varepsilon$ are typically sufficient.

\subsection{Numerical results}

In this section, we present the results obtained from $10^5$ Monte Carlo (MC) replications of the two-signal experiment, considering both the case where the two expected signal counts are treated as parameters of interest and the case where one is tested while the other is treated as a nuisance parameter. In both analyses, the datasets are sampled under the null hypothesis, i.e.\ without signal, with $\beta_0 = 100$, $\phi_1^{(0)} = -0.1, \phi_2^{(0)} = 0.001$ for an energy range $E \in [0.01, 10]$. The peaks' locations have been set to $\gamma_1 = 4$ and $\gamma_2 = 6$ and the exposure to $T = 10$. 

\subsubsection{Two peaks of interest, without nuisance parameters}

We begin by estimating only the two expected signal counts, while the background normalization and shape are treated as fully known, with no nuisance parameters included in the estimation. 
Figure \ref{fig:no_nuis} (left panel) displays the empirical cumulative distribution function (cdf) of $\lrs$ (solid black) and compares it to four reference distributions: (i) the Wilks' regular case, a $\chi^2$ with 2 degrees of freedom (orange dashed), which is visibly too slow near the origin and too light-tailed for this boundary problem; (ii) the Chernoff's distribution, $\frac{1}{2} \chi_0^2 + \frac{1}{2} \chi_1^2$ (teal dotted), which accounts for the presence of one peak on the boundary, but not two; (iii) the chi-squared mixture for two parameters on the boundary provided in \cite{SL} under the non-exposure parametrization that violates Bartlett's identities, indicated as \texttt{mix\_beta} (purple long-dashed), which improves the match but is not fully satisfactory; (iv) the same chi-squared mixture, with the exposure-based parametrization (light-green dash-dot), indicated as \texttt{mix\_exp}, which tracks the empirical cdf most closely across the range. The inset table reports the $50^{\mbox{\footnotesize th}}$, $95^{\mbox{\footnotesize th}}$, and $99^{\mbox{\footnotesize th}}$ quantiles for each curve; the entries closest to the empirical quantiles are emphasized in bold. To ease comparison, the right panel of the figure presents the residual plot $F_{\mbox{\footnotesize ref}} (t) - F_{\mbox{\footnotesize emp}} (t)$, i.e.,~the pointwise difference between each reference cdf and the empirical cdf: positive values indicate a lighter tail than observed, risking overly small $p$-values and hence optimistic significances, while negative values indicate a heavier tail (conservative results). 

{
\setlength{\abovecaptionskip}{-15pt}  % default is often ~10pt
\setlength{\belowcaptionskip}{0pt}  % optional: reduce space below
\begin{figure}[H]
  \centering
  \begin{adjustbox}{width=1.1\textwidth,center}
    \begin{minipage}{0.5\textwidth}
      \centering
      \includegraphics[width=\linewidth]{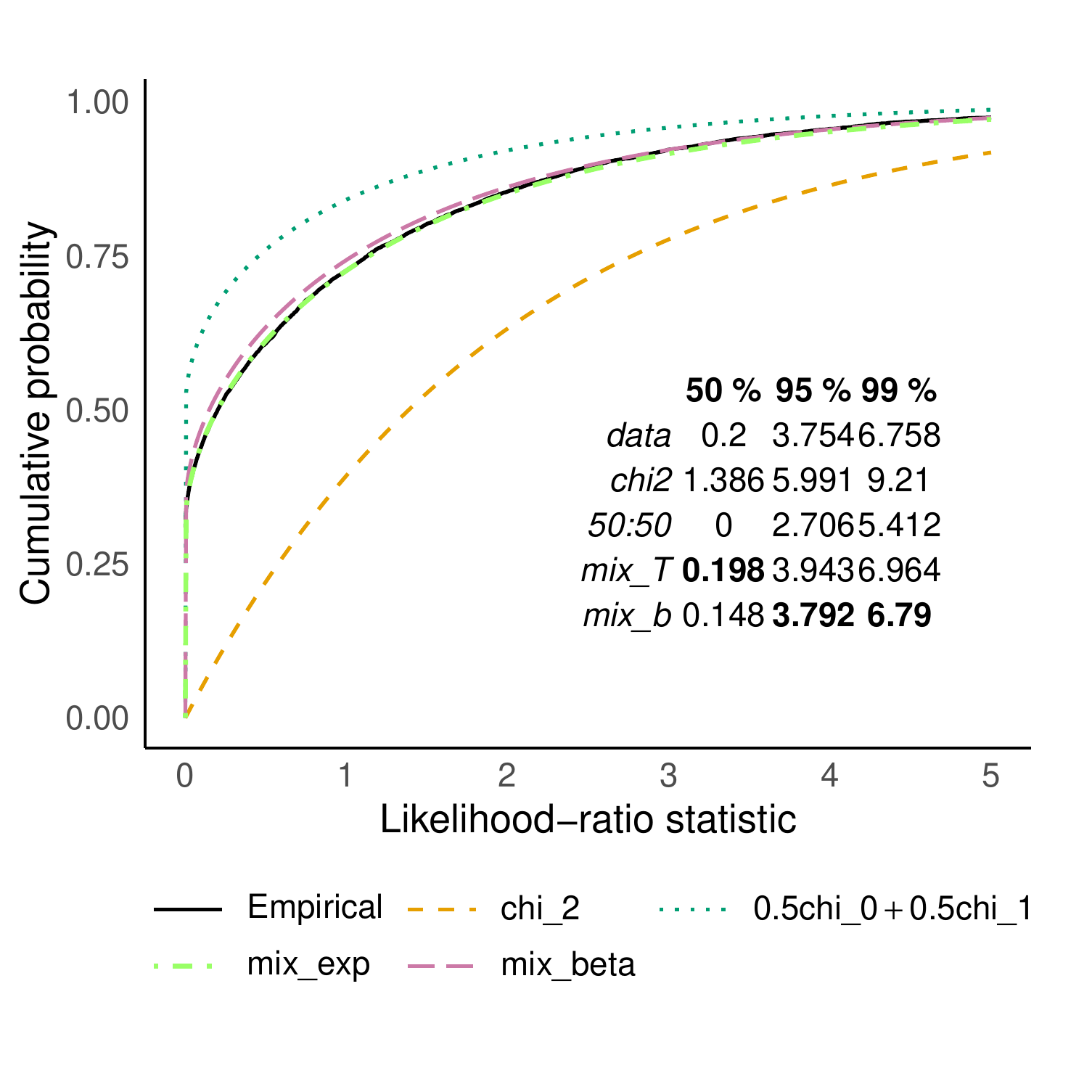}
    \end{minipage}\hfill
    \begin{minipage}{0.5\textwidth}
      \centering
      \includegraphics[width=\linewidth]{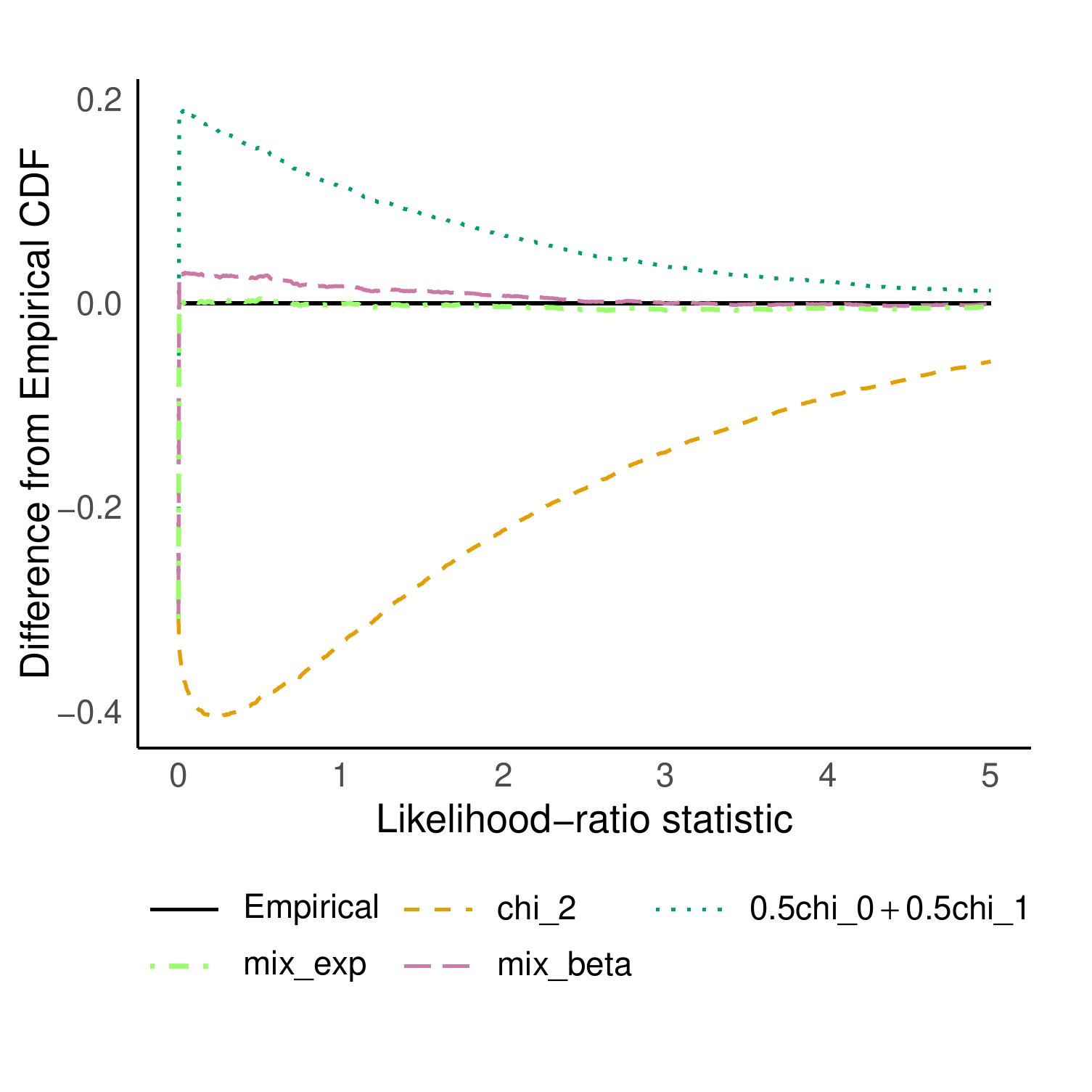}
    \end{minipage}
  \end{adjustbox}
  \caption{\small Diagnostic for the estimation of the two expected signal counts without nuisance parameters. Left: Empirical cumulative distribution function (cdf) of $\lrs$ (solid black), overlaid with four reference cdfs: (i) the Wilks' regular case, a $\chi^2$ with 2 degrees of freedom (orange dashed); (ii) the Chernoff's one-parameter boundary law $\frac{1}{2} \chi_0^2 + \frac{1}{2} \chi_1^2$ (teal dotted); (iii) the chi-squared mixture limit for two parameters on the boundary from \cite{SL}, under the non-exposure parametrization (purple long-dashed); (iv) the same chi-squared mixture, with the exposure-based parametrization (light-green dash-dot). The inset table reports the $50^{\mbox{\footnotesize th}}$, $95^{\mbox{\footnotesize th}}$, and $99^{\mbox{\footnotesize th}}$ quantiles for each curve, with bold highlighting the closest entries to the empirical quantiles. Right: residual plot $F_{\mbox{\footnotesize ref}} (t) - F_{\mbox{\footnotesize emp}} (t)$, where positive values indicate lighter tail than observed, negative values indicate heavier tail.}
  \label{fig:no_nuis}
\end{figure}
}

\subsubsection{Two peaks of interest, with background nuisance parameters}

When the background-related parameters are estimated as well, the situation changes markedly. From Figure~\ref{fig:with_nuis}, we see that neither the \texttt{mix\_beta} nor the \texttt{mix\_exp} curve agrees any longer with the empirical cdf obtained from the MC repetitions. This mismatch arises because, upon profiling out the nuisance parameters, the information matrix must be adjusted accordingly. To make this explicit, let us partition the parameter vector as $\theta = (\psi,\eta)$, where $\psi = (\mu_1,\mu_2)$ are the parameters of interest and $\eta = (\beta,\phi_1,\phi_2)$ denote the nuisance parameters. With this notation, the Fisher information matrix takes the block form

{
\setlength{\abovecaptionskip}{-15pt}  % default is often ~10pt
\setlength{\belowcaptionskip}{0pt}  % optional: reduce space below
\begin{figure}[H]
  \centering
  \begin{adjustbox}{width=1.1\textwidth,center}
    \begin{minipage}{0.5\textwidth}
      \centering
      \includegraphics[width=\linewidth]{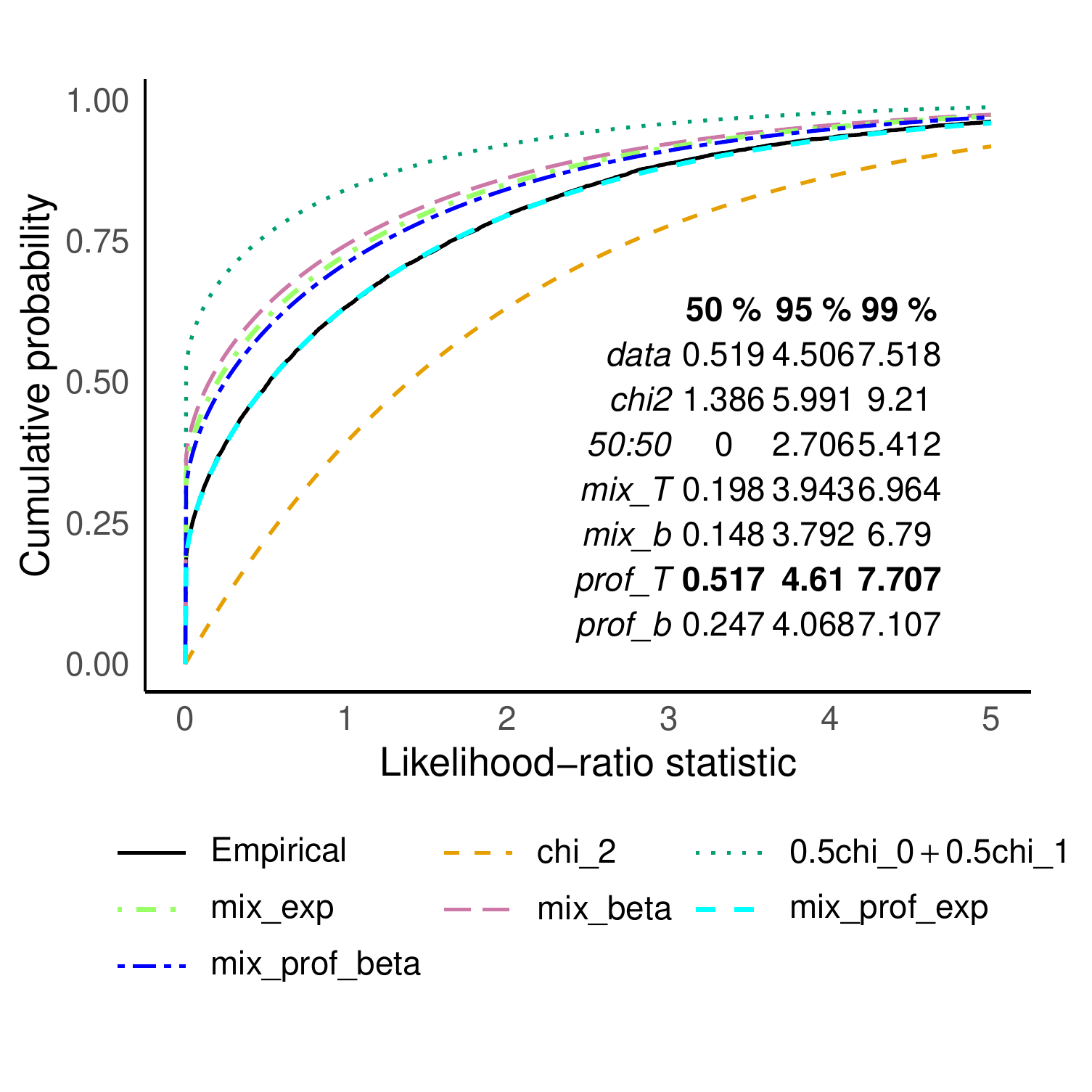}
      \label{fig:left}
    \end{minipage}\hfill
    \begin{minipage}{0.5\textwidth}
      \centering
      \includegraphics[width=\linewidth]{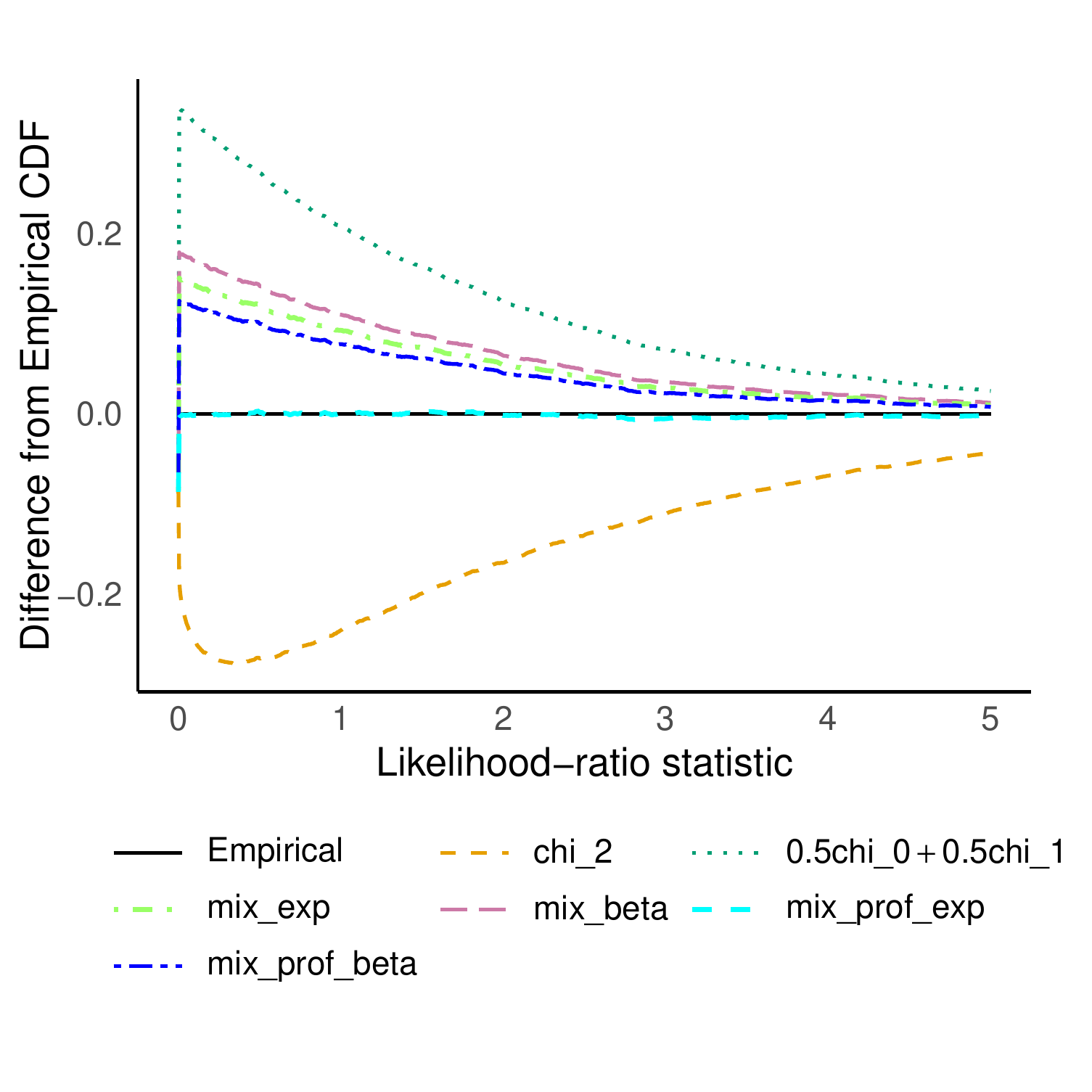}
      \label{fig:right}
    \end{minipage}
  \end{adjustbox}
  \caption{\small Diagnostic for the estimation of the two expected signal counts and the background nuisance parameters. Left: Empirical cumulative distribution function (cdf) of $\lrs$ (solid black), overlaid with six reference cdfs: (i) the Wilks' regular case, a $\chi^2$ with 2 degrees of freedom (orange dashed); (ii) the Chernoff's one-parameter boundary law $\frac{1}{2} \chi_0^2 + \frac{1}{2} \chi_1^2$ (teal dotted); (iii) the chi-squared mixture limit for two parameters on the boundary from \cite{SL}, under the non-exposure parametrization (purple long-dashed); (iv) the same chi-squared mixture, with the exposure-based parametrization (light-green dash-dot); (v) the exposure-based profile chi-squared mixture (cyan long-dash); (vi) the single-event/no-exposure profile chi-squared mixture (dark-blue dash–dot). The inset table reports the $50^{\mbox{\footnotesize th}}$, $95^{\mbox{\footnotesize th}}$, and $99^{\mbox{\footnotesize th}}$ quantiles for each curve, with bold highlighting the closest entries to the empirical quantiles. Right: residual plot $F_{\mbox{\footnotesize ref}} (t) - F_{\mbox{\footnotesize emp}} (t)$, where positive values indicate lighter tail than observed, negative values indicate heavier tail.}
  \label{fig:with_nuis}
\end{figure}
}

\[
\mathcal{I}(\theta) =
\begin{pmatrix}
\mathcal{I}_{\psi\psi} & \mathcal{I}_{\psi\eta} \\
\mathcal{I}_{\eta\psi} & \mathcal{I}_{\eta\eta}
\end{pmatrix},
\]

\noindent where $I_{\psi\psi}$ quantifies information on $(\mu_1,\mu_2)$ if $\eta$ were known, $I_{\eta\eta}$ does so for the nuisance parameters, and $I_{\psi\eta}$ captures their relationship (with $I_{\eta\psi} = I^\top_{\psi\eta} $). The explicit expressions for these blocks for this example are provided in Appendix~B. The appropriate asymptotic reference for inference on $\psi$ is then governed by the \emph{effective} Fisher information obtained via the Schur complement (see, e.g., \cite{pace-salvan}, Section 4.6),

\begin{equation}
\mathcal{I}^\mathrm{eff}_{\psi\psi}
= \mathcal{I}_{\psi\psi}
- \mathcal{I}_{\psi\eta} \,\mathcal{I}_{\eta\eta}^{-1} \,\mathcal{I}_{\eta\psi},
\label{schur}
\end{equation}

\noindent rather than by the unadjusted block $I_{\psi\psi}$ which we have used for the results shown previously. When this correction is implemented, the theoretical curves recover agreement with the simulations, but only under the exposure-based parametrization (cyan long-dash curve in Figure~\ref{fig:with_nuis}, indicated as \texttt{mix\_prof\_exp} in the legend and \texttt{mix\_T} in the inset table): the standard parametrization continues to produce a clear misfit (dark-blue dash–dot curve, indicated as \texttt{mix\_prof\_beta} in the legend and \texttt{prof\_b} in the table), reinforcing the conclusion that the exposure-based parametrization is the appropriate choice.

\subsubsection{First peak of interest, second peak as nuisance, without background nuisance parameters}

Here we consider only the expected signal count of the first peak as the parameter of interest, while the second is treated as a nuisance parameter, still constrained to be positive. This is a common practice in applied scenarios (e.g.\ Higgs-like resonance searches or dark-matter gamma-ray line analyses), where one typically tests one peak at a time: either the second peak is fixed to zero, or it is estimated as a nuisance parameter subject to the physical constraint $\mu_2 \geq 0$. We begin by focusing solely on the strength of the second peak as a nuisance parameter, while keeping all background-related parameters fixed and known. For comparison, in the unphysical but mathematically regular case where the nuisance parameter $\mu_2$ is treated as unconstrained, the likelihood-ratio test statistic obeys the standard Chernoff distribution associated with a single parameter on the boundary.

As described in Section~\ref{sec:lrt}, in the case of a constrained nuisance parameter, the distribution of $\lrs$ depends on the sign of the correlation between $\mu_1$ and $\mu_2$, $\rho_{12} = - \tfrac{I_{12}}{\sqrt{I_{11} I_{22}}}$. If this correlation is positive, the asymptotic distribution is a mixture of $\chi^2_0$, $\chi^2_1$, and $\chi^2_2$ with weights given in Eq.~(\ref{weights_1001}); otherwise, it is no longer a chi-squared mixture, but the correction provided in \cite{us} can be used as an approximation. Since the two signal expected counts are competing parameters, in the sense that an observed excess of events can be attributed either to the first or to the second peak, we expect them to be negatively correlated, with a correlation that increases, in absolute value, as the locations of the two peaks become closer, and as the peaks themselves become broader (with larger standard deviation). In fact, with locations $\gamma_1 = 4$ and $\gamma_2=6$ and standard deviation $\sigma = 1$ as in the present example, we obtain $\rho_{12} = -0.37$. Hence, we expect the empirical cdf from the Monte Carlo study to be in good agreement with the approximation suggested in \cite{us}, which is indeed confirmed by the results shown in Figure~\ref{fig:1001_no_nuis}. It is important to emphasise that the Chernoff distribution, which is typically employed in this type of per-peak testing, is not an appropriate choice, as it fails to account for the fact that, even when treated as a nuisance parameter, the strength of the second peak still lies on the boundary of the parameter space.

\subsubsection{First peak of interest, second peak as nuisance, with background nuisance parameters}

Now, we estimate the background-related parameters as nuisances, together with the strength of the second peak $\mu_2$, while $\mu_1$ is the only parameter of interest of the analysis. Naively, since the number of nuisance parameters remains relatively small, one would not expect substantial changes compared with the previous case. However, when we recompute the correlation between $\mu_1$ and $\mu_2$ using the effective Fisher information (Eq.~\ref{schur}), we now obtain $\rho_{12} = 0.55$. This positive correlation has an important interpretation: once the nuisance parameters are profiled out, the two expected signal counts are no longer purely competing for the same excess of events. Instead, the adjustment induced by the background parameters reshapes the effective information matrix in such a way that the estimates of the two peaks tend to move together. From an inferential point of view, this flips the asymptotic distribution of the likelihood ratio statistic from the ``negative-correlation regime'', where no simple chi-squared mixture applies and we resorted to the heuristic extension, to the ``positive-correlation regime'', where the classical chi-squared mixture weights (Eq.~\ref{weights_1001}) provide the correct asymptotic distribution. Figure~\ref{fig:1001_with_nuis} indeed confirms this, as the empirical cdf matches the positive correlation regime prediction remarkably well, while the heuristic extension, which was the closest to the empirical cdf in the previous case, now visibly fails.

{
\setlength{\abovecaptionskip}{-15pt}  % default is often ~10pt
\setlength{\belowcaptionskip}{0pt}  % optional: reduce space below
\begin{figure}[H]
  \centering
  \begin{adjustbox}{width=1.1\textwidth,center}
    \begin{minipage}{0.5\textwidth}
      \centering
      \includegraphics[width=\linewidth]{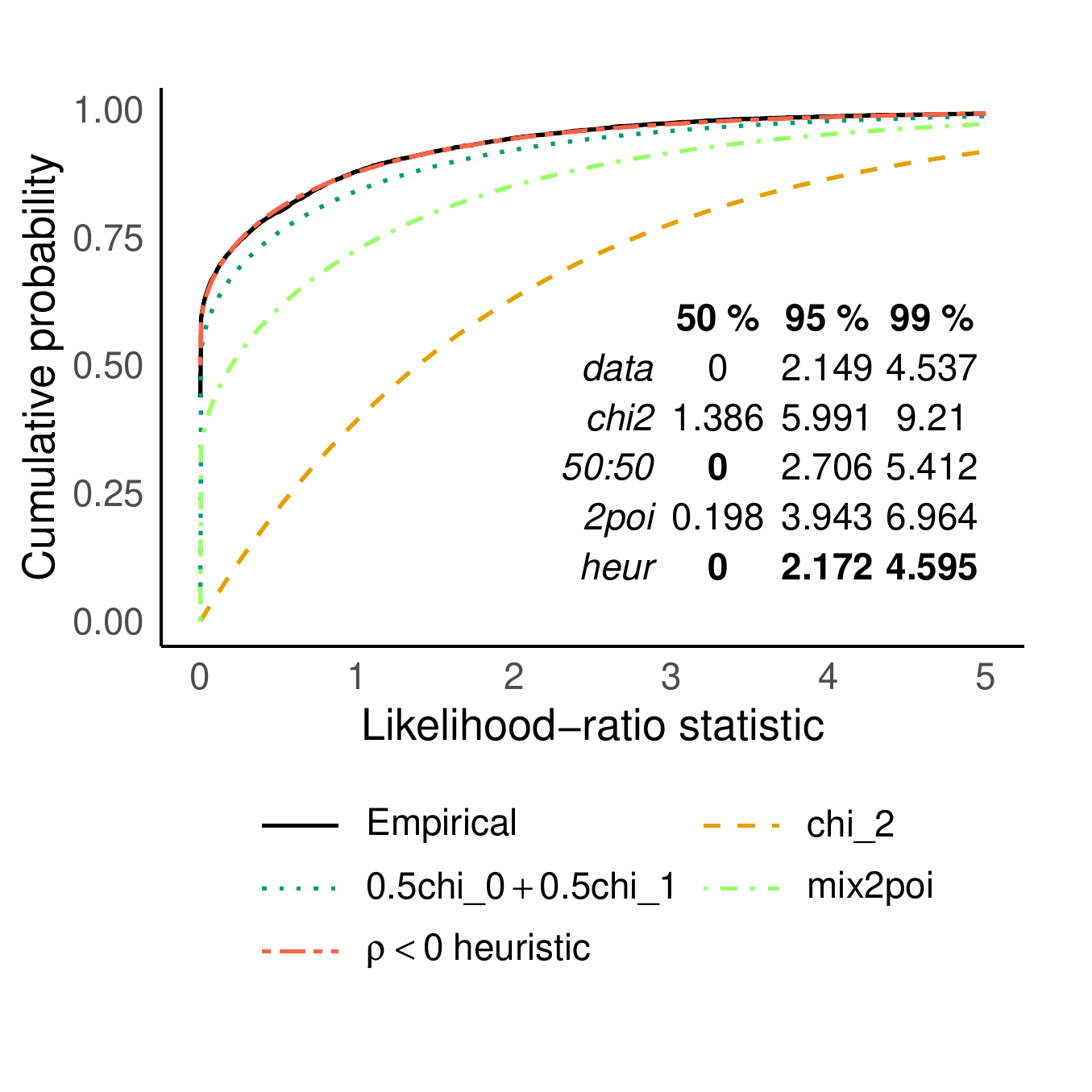}
    \end{minipage}\hfill
    \begin{minipage}{0.5\textwidth}
      \centering
      \includegraphics[width=\linewidth]{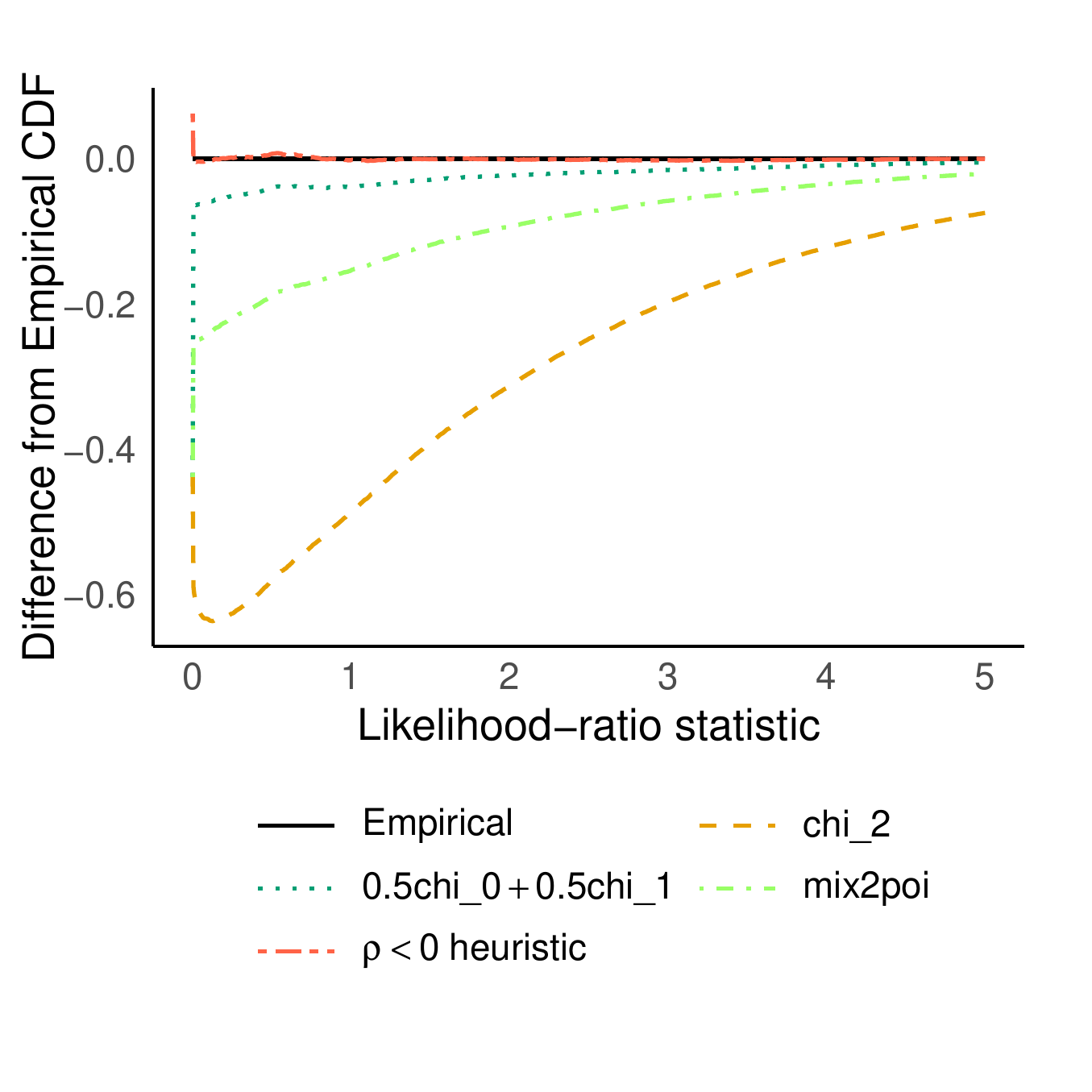}
    \end{minipage}
  \end{adjustbox}
  \caption{\small 
  Diagnostic for the estimation of the first peak yield with the second as nuisance parameter and no additional nuisance from the background. Left: Empirical cumulative distribution function (cdf) of $\lrs$ (solid black), overlaid with four reference cdfs: (i) the Wilks' regular case, a $\chi^2$ with 2 degrees of freedom (orange dashed); (ii) the Chernoff's one-parameter boundary law $\frac{1}{2} \chi_0^2 + \frac{1}{2} \chi_1^2$ (teal dotted); (iii) the chi-squared mixture limit for two parameters on the boundary from \cite{SL}, under the non-exposure parametrization (light-green dash-dot); (iv) the modified distribution for negatively correlated parameters from \cite{us} (red double-dash). The inset table reports the $50^{\mbox{\footnotesize th}}$, $95^{\mbox{\footnotesize th}}$, and $99^{\mbox{\footnotesize th}}$ quantiles for each curve, with bold highlighting the closest entries to the empirical quantiles. Right: residual plot $F_{\mbox{\footnotesize ref}} (t) - F_{\mbox{\footnotesize emp}} (t)$, where positive values indicate lighter tail than observed, negative values indicate heavier tail.}
  \label{fig:1001_no_nuis}
\end{figure}
}

\section{Discussion} \label{sec:discussion}

This analysis provides a detailed characterization of the distribution of the likelihood ratio statistic $\lrs$ in background–plus–signal settings where two parameters are constrained to the boundary of the parameter space.

In the simplest setting, with the expected count from two peaks as parameters of interest and no background nuisance parameters, the empirical distribution of $\lrs$ clearly departs from Wilks' $\chi^2_2$ law and from the single-parameter Chernoff distribution. The chi-squared mixture for two parameters on the boundary \cite{SL}, yields some improvement when applied within the standard extended-likelihood parametrization. However, only the exposure-based parametrization achieves close agreement with the empirical distribution across the full range: this corroborates the appropriateness of the exposure parametrization when formulating the test, as it restores Bartlett's identities and allows the use of standard asymptotic theory.

{
\setlength{\abovecaptionskip}{-15pt}  % default is often ~10pt
\setlength{\belowcaptionskip}{0pt}  % optional: reduce space below
\begin{figure}[H]
  \centering
  \begin{adjustbox}{width=1.1\textwidth,center}
    \begin{minipage}{0.5\textwidth}
      \centering
      \includegraphics[width=\linewidth]{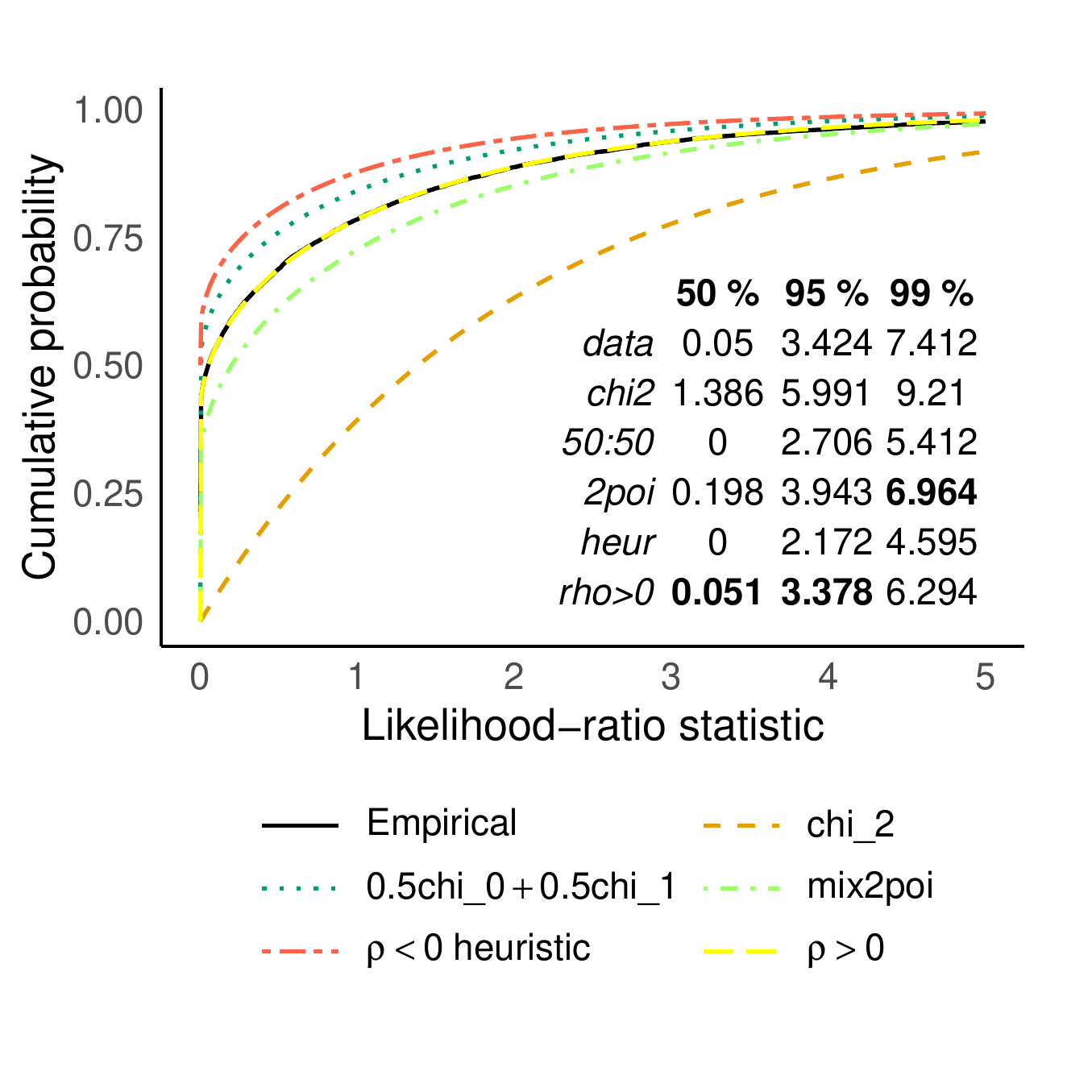}
    \end{minipage}\hfill
    \begin{minipage}{0.5\textwidth}
      \centering
      \includegraphics[width=\linewidth]{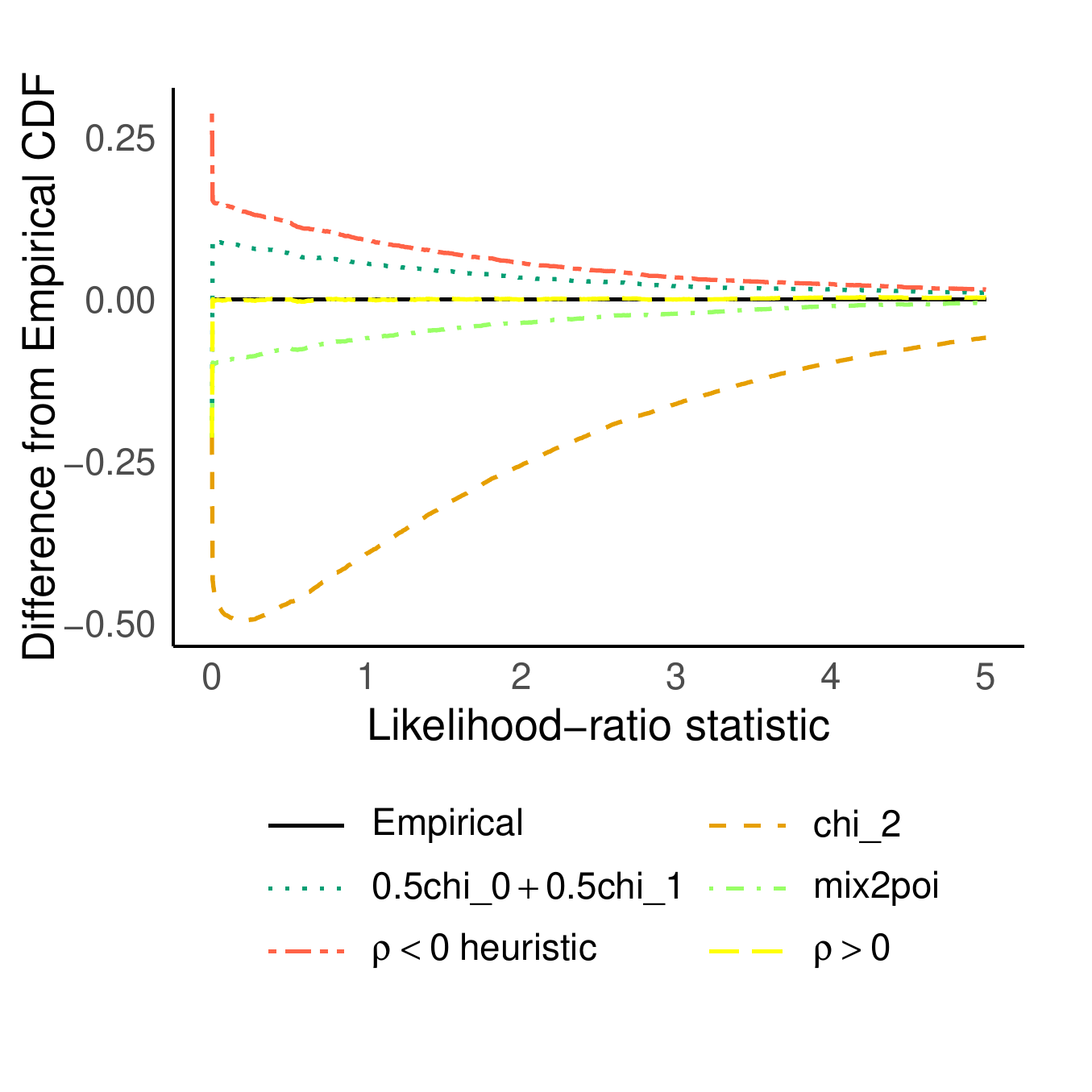}
    \end{minipage}
  \end{adjustbox}
  \caption{\small Diagnostic for the estimation of the first peak yield with the second as nuisance parameter and the three additional nuisance from the background. Left: Empirical cumulative distribution function (cdf) of $\lrs$ (solid black), overlaid with five reference cdfs: (i) the Wilks' regular case, a $\chi^2$ with 2 degrees of freedom (orange dashed); (ii) the Chernoff's one-parameter boundary law $\frac{1}{2} \chi_0^2 + \frac{1}{2} \chi_1^2$ (teal dotted); (iii) the chi-squared mixture limit for two parameters on the boundary from \cite{SL}, under the non-exposure parametrization (light-green dash-dot); (iv) the modified distribution for negatively correlated parameters from \cite{us} (red double-dash); (v) the exact chi-squared mixture for the positive correlation case (yellow long-dashed). The inset table reports the $50^{\mbox{\footnotesize th}}$, $95^{\mbox{\footnotesize th}}$, and $99^{\mbox{\footnotesize th}}$ quantiles for each curve, with bold highlighting the closest entries to the empirical quantiles. Right: residual plot $F_{\mbox{\footnotesize ref}} (t) - F_{\mbox{\footnotesize emp}} (t)$, where positive values indicate lighter tail than observed, negative values indicate heavier tail.}
  \label{fig:1001_with_nuis}
\end{figure}
}

When background nuisance parameters are introduced, the same mixture distribution ceases to match the simulations, illustrating that the effective Fisher information must be used in place of the unadjusted block to account for the effect of profiling. Implementing the Schur complement correction restores agreement, but, notably, only under the exposure-based parametrization. This confirms that the exposure-based parametrization is essential for correct asymptotics in the presence of nuisance parameters.

Often, to avoid dealing with a double-peak simultaneous test, the literature presents per-peak tests where the second peak is regarded as a nuisance parameter. Turning to such a per-peak testing framework, we first considered the case where $\mu_1$ is the parameter of interest and $\mu_2$ is treated as a nuisance, with all background parameters fixed. Here the two peaks act as competing explanations for the same excess, leading to a negative correlation. In this regime, the correct asymptotic distribution is no longer a simple chi-squared mixture, and the empirical distribution is best approximated by the heuristic correction proposed in \cite{us}. Importantly, the Chernoff distribution $\tfrac{1}{2}\chi^2_0 + \tfrac{1}{2}\chi^2_1$, which is often used in applied per-peak testing, is inadequate: even as a nuisance, the second peak remains constrained to the boundary, and this structural feature must be accounted for.

Finally, when both the background nuisance parameters and $\mu_2$ are profiled, the correlation flips sign ($\rho_{12}>0$): in this example, profiling reshapes the information structure so that the two expected counts tend to move together rather than compete. This transition from negative to positive correlation must be taken into account, as it shifts the reference distribution from the non-$\chi^2$ distribution approximated with the heuristic modification to the chi-squared mixture regime with weights determined by the effective Fisher information as per Eq.(\ref{weights_1001}). This illustrates that even modest nuisance sets can qualitatively change the regime, hence the effective Fisher information must be checked explicitly to use the correct asymptotic distribution.

These findings have direct parallels in applied analyses. In independent double-peaked gamma-ray line searches with Fermi–LAT data \cite{FermiLAT, anderson_2016, rev_2}, the two-line yields are left as free parameters, hence with a null distribution corresponding to our two-peak setup. Such a situation requires a chi-squared mixture with weights determined by the profiled Fisher information. Nevertheless, Wilks-based mappings with clipping,  asymptotically equivalent to the Chernoff distribution, are still often applied; in some cases, a plain $\chi^2_2$ distribution is even used \cite{rev_2}, an assumption demonstrably inconsistent with the boundary problem at hand. As illustrated by our results, adopting the Chernoff distribution results in overly optimistic significance estimates, whereas employing the plain chi-squared distribution from Wilks' theorem leads to conservative significance estimates. Beyond gamma-ray lines, similar practices occur in collider resonance spectroscopy, where one typically tests one peak at a time or compares one-peak vs two-peak hypotheses within extended-likelihood fits to invariant-mass spectra \cite{cms, lhcb}. A concrete example is provided by extended Higgs sectors, where a CP-even/CP-odd H/A pair can generate two neighbouring peaks in a common final state \cite{higgs}, for which a joint two-signal analysis is well suited. Our results emphasize that whenever the strength of a second peak is treated as a nuisance and constrained to be positive, the effective correlation induced by profiling the background determines which asymptotic reference is correct. 

More generally, the framework discussed in this paper is broadly applicable across several experimental contexts. In helioscope and X-ray searches for axions, theory can yield two discrete lines within the same band \cite{helioscope}, motivating a two-line joint test. In kinematic-feature searches (double endpoints or twin Jacobian peaks from cascade decays), two sharp structures predicted by a mass hierarchy \cite{jacob} can be tested together. In paired signatures across channels (for example, a resonance contributing to both $\gamma\gamma$ and $Z\gamma$ spectra \cite{paired_signature}), a single combined extended likelihood with two signal strengths cleanly captures the model relation. 

In all these settings and potentially many more, when testing two non-negative counts, the appropriate reference is the chi-squared mixture with weights determined by the profiled Fisher information described here. In fact, in Figure~\ref{fig:sigma} we provide a direct comparison between the tails of the most commonly used reference distributions and the correct mixture, together with the corresponding $3$--$5\sigma$ levels. 
The upper panel corresponds to the two-peak test, while the lower panel shows the per-peak test where the second peak is treated as a nuisance. In both  settings, using the one-parameter Chernoff distribution assigns too small a test-statistic value to a given $\sigma$ level, thereby inflating the significance and risking false discovery claims. Conversely, using the two-parameter Wilks distribution assigns thresholds that are too large, making the test overly conservative and potentially missing real discoveries.

\section{Conclusion} \label{sec:conclusion}

In this work we characterized the asymptotic distribution of the likelihood ratio test statistic when two parameters of interest lie on the boundary of the parameter space, using a double-peak counting experiment as an example to illustrate the effect of nuisance parameters, parametrization choices, and correlation structure. Our simulations show that na\"ive use of Wilks' theorem or the Chernoff distribution can lead to severe misspecification, and that the correct asymptotic reference depends critically on the parametrization and on the effect of profiling nuisance parameters. Using an overly conservative asymptotic distribution (heavier-tailed) leads to $p$-values that are larger than their true values, which may cause genuine discoveries to be missed. Conversely, adopting an overly liberal asymptotic distribution (lighter-tailed) yields artificially small  $p$-values, increasing the risk of wrongful discovery claims. 

For the present case, exposure-based parametrization guarantees the validity of asymptotic theory when enough data are observed, and allows the use of chi-squared mixtures with weights determined by the profiled Fisher information. The results highlight that even modest sets of nuisance parameters can qualitatively alter the correlation structure between boundary parameters and thereby change the limiting distribution. Correctly diagnosing the regime by examining the effective Fisher information is essential for valid inference.

{
\setlength{\abovecaptionskip}{10pt}  % default is often ~10pt
\setlength{\belowcaptionskip}{0pt}  % optional: reduce space below
\begin{figure}[H]
      \centering
      \includegraphics[width=\linewidth]{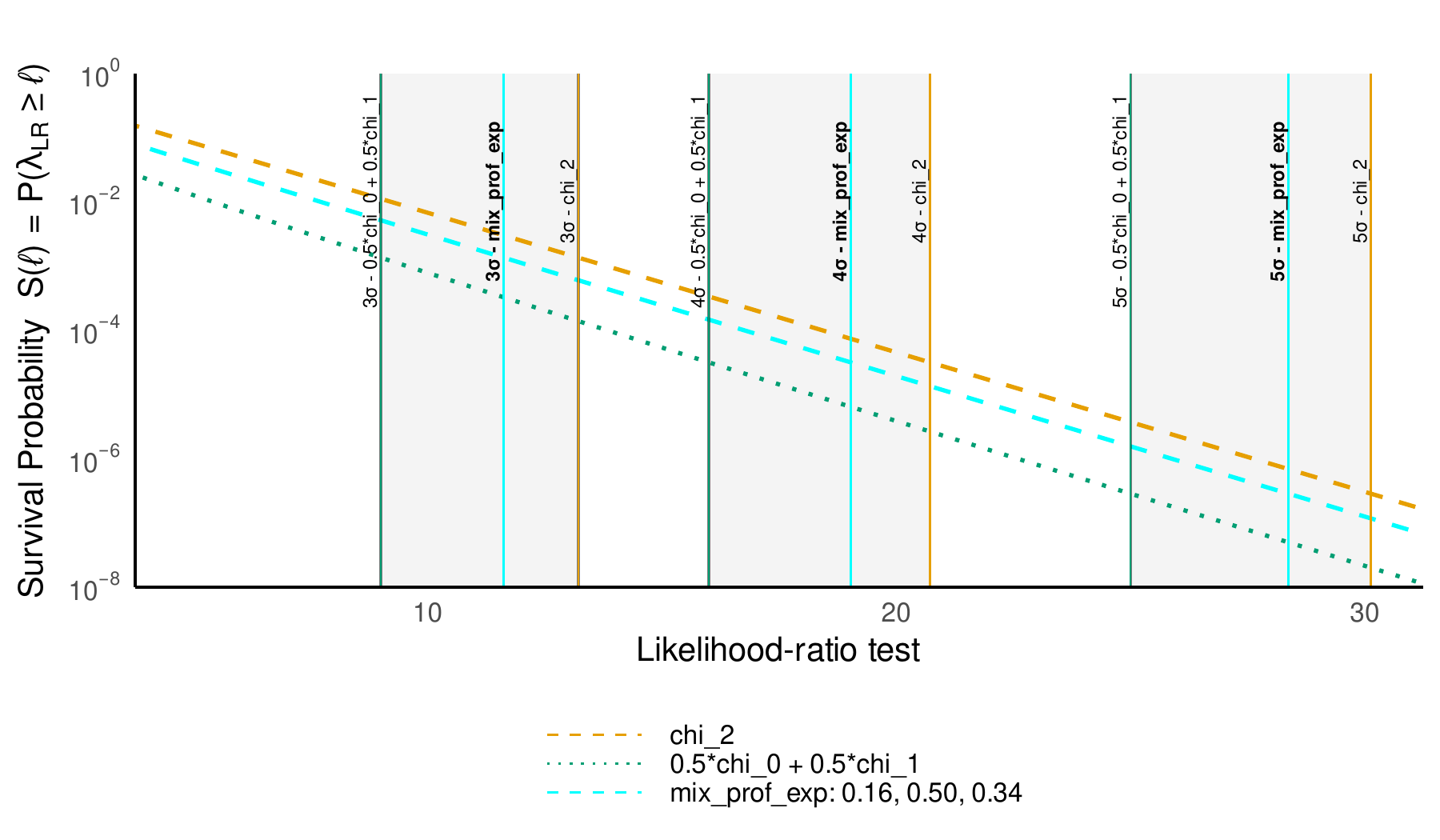}
      \includegraphics[width=\linewidth]{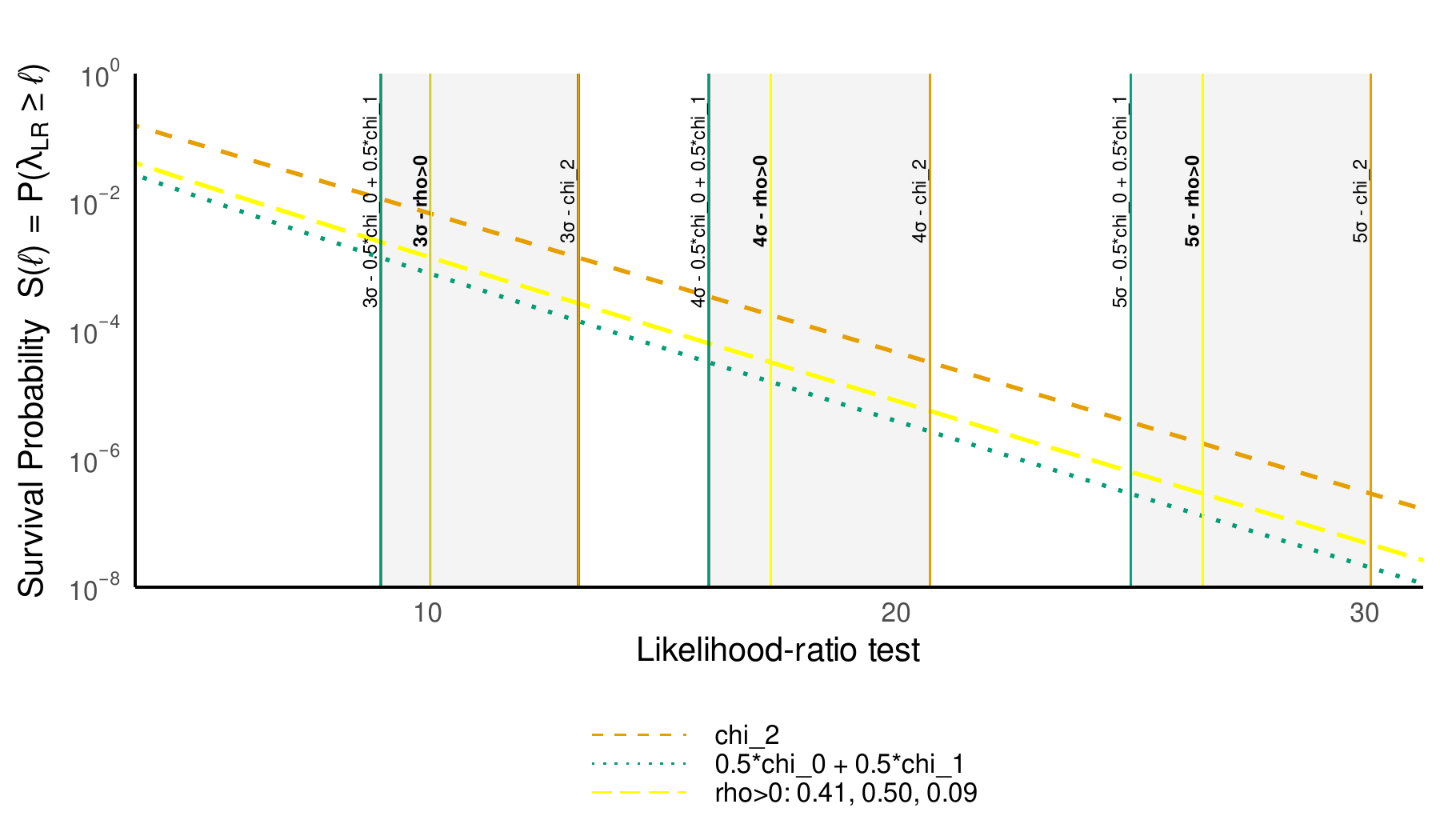}
  \caption{\small Survival probability, i.e., upper tails of the likelihood–ratio distribution under various asymptotic references (log-scale). Upper panel: two–peak test with both expected counts from the signal as parameters of interest. Lower panel: per–peak test where the second peak is treated as a nuisance. In each panel, the
orange dashed curve corresponds to the Wilks $\chi^2_2$ distribution; the green dash–dotted to the Chernoff $\tfrac12\chi^2_0+\tfrac12\chi^2_1$ distribution; and the cyan (upper panel) or yellow (lower panel) mixed solid curve shows the chi–squared mixture with weights determined by the profiled Fisher information. The grey shaded regions indicate the span of the nominal $3\sigma$, $4\sigma$, and $5\sigma$ thresholds implied by the various asymptotic reference models, while the coloured vertical lines mark the exact locations of these thresholds, using matching line colours.
%\anders{Can we maybe put these figures underneath each other and increase the size of each? In the current configuration the different labels are very small compared to the main text in the paper. Also, maybe change the y axis labels from computer notation ``1e+00'' to regular powers of 10?}
}
  \label{fig:sigma}
\end{figure}
}

\section*{Acknowledgments}

This work has been supported by the Research Council of Norway (RCN) through the FRIPRO PLUMBIN' (grant no.~323985), the Centre of Excellence INTEGREAT (no.~332645), and the Norwegian Centre for CERN Research (no.~310713).

\bibliographystyle{unsrt}
\bibliography{refs}

\newpage

\section*{Appendix A} \label{appendix}

In this appendix, we show that under the parametrization of Eq.\ (\ref{likelihood}) Bartlett's identities are not satisfied. For simplicity, and limited to the derivation in this section, we consider the case of a single signal peak with strength $\mu$, normalized density $g(E)$, and null hypothesis $H_0 : \mu = 0$; the results are immediately generalizable to the double-peak case. According to the two Bartlett's identities, under interchangeability of differentiation and integration signs and smoothness in $\theta$, under the null ($\theta = \theta_0)$ we have:

\begin{equation}
    \begin{cases}
        \mathbb{E}[\nabla \ell(\theta)]_{\theta=\theta_0} = 0 \\[1em]
        \mathcal{I}(\theta_0) = -\,\mathbb{E}\bigl[\nabla^2\ell(\theta)\bigr]_{\theta=\theta_0} = \mbox{Var}[\nabla \ell(\theta)]_{\theta=\theta_0}
    \end{cases}
\end{equation}

The first identity states that the score is an unbiased estimating function for $\theta$, while the second equates its covariance to the expected curvature of the log-likelihood; these relationships are a sufficient condition for Wilks' theorem to hold in absence of boundary issues or other irregularities \cite{brazzale}. 
In a regular likelihood setting with iid (independent and identically distributed) observations, both Bartlett's identities are supposed to hold, both at the single observational unit (per-observation likelihood) and at the level of the whole sample of $N$ observations (full likelihood).

When considering the extended maximum likelihood, which does not fully correspond to the standard method of maximum likelihood \cite{barlow}, Bartlett's identities are satisfied when the entire sample of $N$ events is considered. In fact, letting $h(E_i) = \mu g(E_i) + \beta b(E_i)$ for $i = 1,...,N$, the full likelihood can be written as

\[
L(\mu; \beta, \phi_1, \phi_2)
= \frac{e^{-\lambda} \lambda^N}{N!}
  \prod_{i=1}^N \frac{h(E_i)}{\lambda},
\]

\noindent with log-likelihood 
\begin{equation}
    \ell(\mu; \beta, \phi_1, \phi_2)
= -\lambda + \sum_{i=1}^N \log h(E_i) + \text{const}.
\label{loglikelihood}
\end{equation}

\noindent The score function with respect to the signal expected count is
\begin{equation}
\frac{\partial \ell(\mu; \beta, \phi_1, \phi_2)}{\partial \mu}
= -1 + \sum_{i=1}^N \frac{g(E_i)}{h(E_i)}.
\label{score}
\end{equation}

\noindent To compute the expectation of the score, we use that 

\[ 
\mathbb{E}\left[ \sum_{i=1}^N \frac{g(E_i)}{h(E_i)} \right] =  \mathbb{E}\left[ \mathbb{E}\left[ \sum_{i=1}^N \frac{g(E_i)}{h(E_i)}  \biggl \vert N \right] \right] \] 

\noindent and, since  
\[ \mathbb{E}\left[ \sum_{i=1}^N \frac{g(E_i)}{h(E_i)}  \biggl \vert N \right] =  \sum_{i=1}^N \mathbb{E}\left[ \frac{g(E_i)}{h(E_i)}  \biggl \vert N \right] = \sum_{i=1}^N \int_{\EMIN}^{\EMAX} \frac{h(E)}{\lambda} \frac{g(E)}{h(E)} dE = \frac{N}{\lambda}, \]

\noindent we get 

\[
\mathbb{E}[\nabla \ell(\theta)] = -1 + \mathbb{E}\left[ \frac{N}{\lambda} \right] = -1 + \frac{\lambda}{\lambda} = 0,
\]

\noindent therefore the first Bartlett identity is satisfied.

\vspace{\baselineskip}

\noindent Moreover, differentiating the score function, we obtain
\begin{equation}
-\frac{\partial^2 \ell(\mu; \beta, \phi_1, \phi_2)}{\partial \mu^2}
= \sum_{i=1}^N \frac{g(E_i)^2}{h(E_i)^2}.
\label{neg_obs}
\end{equation}

\noindent Taking the expectation,

\begin{align*}
\mathcal{I}_{\mu}(\theta) &= \mathbb{E}\left[ \sum_{i=1}^N \frac{g(E_i)^2}{h(E_i)^2} \right] =  \mathbb{E}\left[ \mathbb{E}\left[ \sum_{i=1}^N \frac{g(E_i)^2}{h(E_i)^2}  \biggl \vert N \right] \right] \\ &= \mathbb{E}\left[  N \int_{\EMIN}^{\EMAX} \frac{1}{\lambda} \frac{g(E)^2}{h(E)} dE \right] = 
  \int_{\EMIN}^{\EMAX} \frac{g(E)^2}{h(E)} dE
\end{align*}

\noindent which under the null simplifies to

\begin{equation}
\mathcal{I}_{\mu}(\theta_0) =  \frac{1}{\beta_0}\int_{\EMIN}^{\EMAX} \frac{g(E)^2}{b(E)} dE,
\label{res_whole_sample}
\end{equation}

\noindent where $\beta_0$ denotes the true value of the expected background count $\beta$.

\noindent This expression has to be compared with the variance of the score, which we can write, by means of the law of the total variance, as

\begin{equation}
    \mbox{Var}[\nabla \ell(\theta)] = \mathbb{E} \left[ \mbox{Var}[\nabla \ell(\theta) \, \vert \, N] \right] + \mbox{Var} \left[ \mathbb{E}[\nabla \ell(\theta) \, \vert \, N] \right] 
\end{equation}

\noindent with, as seen before, $\mathbb{E}[\nabla \ell(\theta) \, \vert \, N] = -1 + N/\lambda$, and with $\mbox{Var}[\nabla \ell(\theta) \, \vert \, N] = N \, \mbox{Var}[g(E_i)/h(E_i)]$. Then,

\begin{align*}
    \mbox{Var}[\nabla \ell(\theta)] &= \mathbb{E} \left[ N \, \mbox{Var}\left[\frac{g(E_i)}{h(E_i)}\right] \right] + \mbox{Var} \left[ -1 + \frac{N}{\lambda} \right] \\[1em]
    &= \lambda \, \mbox{Var}\left[\frac{g(E_i)}{h(E_i)}\right] + \frac{1}{\lambda} \\[1em]
    &= \lambda \left( \mathbb{E}\left[\frac{g(E_i)^2}{h(E_i)^2}\right] - \mathbb{E}\left[\frac{g(E_i)}{h(E_i)}\right]^2 \right) + \frac{1}{\lambda} \\[1em]
    &= \lambda \left( \frac{1}{\lambda} \int_{\EMIN}^{\EMAX} \frac{g(E)^2}{h(E)} dE - \frac{1}{\lambda^2}\right) + \frac{1}{\lambda} \\[1em]
    &=  \int_{\EMIN}^{\EMAX} \frac{g(E)^2}{h(E)} dE 
\end{align*}

\noindent which under the null becomes $\mbox{Var}[\nabla \ell(\theta)]_{\theta = \theta_0} = \frac{1}{\beta_0} \int_{\EMIN}^{\EMAX} \frac{g(E)^2}{b(E)} dE$ and corresponds to the Fisher information formula in Eq.\ (\ref{res_whole_sample}), confirming that, when considering the full likelihood, also the second Bartlett identity holds.

\vspace{\baselineskip}

\noindent However, this no longer holds if the single observed event is considered as observational unit. In fact, the likelihood of Eq.\ (\ref{likelihood}) for the $i\text{-th}$ observation becomes

\[
L_i(\theta) = \lambda \, e^{-\lambda} \, \frac{1}{\lambda} \, h(E_i) \quad \quad \mbox{with } \;
h(E_i) = \mu \, g(E_i) + \beta \, b(E_i; \phi_1, \phi_2), 
\]

\noindent with log-likelihood

\[
\ell_i(\theta) = -\lambda + \log h(E_i).
\]

\noindent The score is 

\[ \frac{\partial \ell_i(\theta)}{\partial \mu}
= -1 + \frac{g(E_i)}{h(E_i)} \] 

\noindent hence 

\[
\mathbb{E} \left[\frac{\partial \ell_i(\theta)}{\partial \mu}\right] = -1 + \frac{1}{\lambda} \qquad \mbox{and} \qquad \mathbb{E} \left[\frac{\partial \ell_i(\theta)}{\partial \mu}\right]_{\theta = \theta_0} = -1 + \frac{1}{\beta_0}
\]

\noindent which is not null unless $\beta_0 = 1$. Therefore, the first Bartlett identity is not generally guaranteed to be valid.

\vspace{\baselineskip}

Taking the expectation of the negative second derivative, we get

\begin{equation}
    \mathbb{E} \left[- \frac{\partial^2 \ell_i(\theta)}{\partial \mu^2} \right] = \mathbb{E} \left[ \frac{g(E_i)^2}{h(E_i)^2} \right] = \frac{1}{\lambda} \int_{\EMIN}^{\EMAX} \frac{g(E)^2}{h(E)} dE,
\end{equation}

\noindent hence

\begin{equation}
    \mathbb{E} \left[ \frac{g(E_i)^2}{h(E_i)^2} \right]_{\theta = \theta_0} = \frac{1}{\beta_0^2} \int_{\EMIN}^{\EMAX} \frac{g(E)^2}{b(E)} dE;
\end{equation}

\noindent however, this expression does not equate the variance of the score, 

\[
\mbox{Var} \left( \frac{\partial \ell_i (\theta)}{\partial \mu} \right) =  
\mbox{Var} \left( -1 + \frac{g(E_i)}{h(E_i)} \right) = \frac{1}{\lambda} \int_{\EMIN}^{\EMAX} \frac{g(E)^2}{h(E)} dE - \frac{1}{\lambda^2},
\]

\noindent which under the null becomes 

\[
\mbox{Var} \left[ \frac{\partial \ell_i(\theta)}{\partial \mu} \right]_{\theta = \theta_0} = \frac{1}{\beta_0^2} \int_{\EMIN}^{\EMAX} \frac{g(E)^2}{b(E)} dE - \frac{1}{\beta_0^2},
\]

\noindent therefore the second identity also does not hold.

\vspace{2cm}
\section*{Appendix B} \label{appendixB}

In this appendix, we provide explicit expressions for the expected Fisher information entries for the two-peak discovery example considered in the article. Let the parameter vector be partitioned as $\theta=(\psi,\eta)$ with
\[
\psi=(\mu_1,\mu_2),\qquad \eta=(\beta,\phi_1,\phi_2),
\]
so that the Fisher information takes the block form
\[
I(\theta)\;=\;
\begin{pmatrix}
I_{\psi\psi} & I_{\psi\eta} \\
I_{\eta\psi} & I_{\eta\eta}
\end{pmatrix}.
\]

\noindent With this partition, the $2\times 2$ block $I_{\psi\psi}$ collects the entries
$I_{\mu_i\mu_j}$, while the $2\times 3$ block $I_{\psi\eta}$ collects the entries
$I_{\mu_i\beta}$ and $I_{\mu_i\phi_k}$, and similarly for the remaining blocks. With $h(E)=h(E;\theta)$ and derivatives $\partial_{\phi_k}b=\frac{\partial}{\partial \phi_k}b(E;\phi_1,\phi_2)$, the individual components are given by:
\begin{align*}
I_{\mu_i\mu_j}(\theta)
&= \frac{T}{\lambda}\int_{E_{\min}}^{E_{\max}} \frac{g_i(E)\,g_j(E)}{h(E)}\,dE, 
&& i,j\in\{1,2\};\\[0.6em]
I_{\mu_i\beta}(\theta)
&= \frac{T}{\lambda}\int_{E_{\min}}^{E_{\max}} \frac{g_i(E)\,b(E)}{h(E)}\,dE,
&& i\in\{1,2\};\\[0.6em]
\end{align*}
\begin{align*}
I_{\beta\beta}(\theta)
&= \frac{T}{\lambda}\int_{E_{\min}}^{E_{\max}} \frac{b(E)^2}{h(E)}\,dE;\\[0.6em]
I_{\mu_i\phi_k}(\theta)
&= \frac{T\beta }{\lambda}\int_{E_{\min}}^{E_{\max}} \frac{g_i(E)\,\partial_{\phi_k} b(E)}{h(E)}\,dE,
&& i\in\{1,2\},\; k\in\{0,1\};\\[0.6em]
I_{\beta\phi_k}(\theta)
&= \frac{T\beta}{\lambda}\int_{E_{\min}}^{E_{\max}} \frac{b(E)\,\partial_{\phi_k} b(E)}{h(E)}\,dE,
&& k\in\{0,1\};\\[0.6em]
I_{\phi_k\phi_l}(\theta)
&= \frac{T\beta^2}{\lambda}\int_{E_{\min}}^{E_{\max}} \frac{\partial_{\phi_k} b(E)\,\partial_{\phi_l} b(E)}{h(E)}\,dE,
&& k,l\in\{0,1\},
\end{align*}
where $T$ is the exposure time and all integrals are over $[E_{\min},E_{\max}]$.

\end{document}